\documentclass[american]{aastex}
\usepackage[T1]{fontenc}
\usepackage[latin9]{inputenc}
\usepackage{float}
\usepackage{graphicx}

\makeatletter

\providecommand{\tabularnewline}{\\}

\slugcomment{Submitted to Astrophysical Journal Letters}
\shorttitle{Jumping Jupiter can explain Mercury's orbit}
\shortauthors{Roig et al.}

\makeatother

\usepackage{babel}
\begin{document}

\title{Jumping Jupiter can explain Mercury's orbit}

\author{Fernando Roig}

\affil{Observatório Nacional, Rua Gal. Jose Cristino 77, Rio de Janeiro,
RJ 20921-400, Brazil}

\email{froig@on.br}

\author{David Nesvorný}

\affil{Southwest Research Institute, 1050 Walnut St., Suite 300, Boulder,
CO 80302, USA}

\email{davidn@boulder.swri.edu}

\author{Sandro Ricardo DeSouza}

\affil{Observatório Nacional, Rua Gal. Jose Cristino 77, Rio de Janeiro,
RJ 20921-400, Brazil}

\email{sandroricardo@on.br}
\begin{abstract}
The orbit of Mercury has large values of eccentricity and inclination
that cannot be easily explained if this planet formed on a circular
and coplanar orbit. Here, we study the evolution of Mercury's orbit
during the instability related to the migration of the giant planets
in the framework of the jumping Jupiter model. We found that some
instability models are able to produce the correct values of Mercury's
eccentricity and inclination, provided that relativistic effects are
included in the precession of Mercury's perihelion. The orbital
excitation is driven by the fast change of the normal oscillation
modes of the system corresponding to the perihelion precession of
Jupiter (for the eccentricity), and the nodal regression of Uranus
(for the inclination).
\end{abstract}


\section{Introduction}

The orbit of Mercury is the most peculiar among the Solar System planets.
It has the largest mean eccentricity and inclination ($e=0.17$ and
$I=7^{\circ}$ with respect to the invariable plane), its perihelion
longitude $\varpi$ is significantly affected by relativistic effects
($\delta\dot{\varpi}\sim0.43\arcsec\,\mathrm{yr}^{-1}$), and its
orbit displays the most chaotic long term dynamics \citep{1994A&A...287L...9L}.
Secular variations have typical amplitudes of $\Delta e\simeq\pm0.08$
and $\Delta I\simeq\pm3^{\circ}$ that are not large enough to explain
the current mean values of $e$ and $I$ assuming that Mercury, as
well as the other terrestrial planets, ended the formation process
on nearly circular and coplanar orbits. 

\citet{1976Icar...28..441W} proposed that the high $e,I$ of Mercury
could be produced by the oblateness perturbation of the Sun. An initially
large value of the second degree harmonic $J_{2}$ caused by a rapidly
rotating Sun and its subsequent spin down, would drive the perihelion
and node of Mercury into secular resonances with the perihelion and
node of Venus, respectively, that would excite the corresponding proper
modes of Mercury's orbit. However, recent observational evidence
indicate that, by the time the terrestrial planets are in their final
stages of formation ($\sim30$ to 200~Myr depending on the model),
the Sun's rotation would be only a few times faster than the current
rotation (\citealp{2013EAS....62..143B}). Moreover, contraction of
the star along the pre-main-sequence track occurs on the Kelvin-Helmholtz
timescale ($\sim30$~Myr for the Sun), implying that the $J_{2}$
term would be already small when the terrestrial planets are formed.

\citet{2008Icar..196....1L}, \citet{2009Natur.459..817L}, and \citet{2012A&A...548A..43B}
showed that the chaotic evolution of the orbit of Mercury may lead
to large changes of $e,I$ over the age of the Solar System. According
to \citet{2008Icar..196....1L}, the probability of changing Mercury's
eccentricity by more than 0.1 over 5~Gyr of evolution is less than
20\%. The probability of changing the inclination by more than $5^{\circ}$
over the same time scale is even smaller, less than 10\%. Therefore,
typically the chaotic variations of Mercury's orbit do not appear
to be able to provide the required excitation in $e$ and $I$ from
initially circular and coplanar orbits. In particular, \citet{2009Natur.459..817L}
performed about 2500 simulations and in none of them was Mercury able
to reach an eccentricity smaller than $\sim0.05$ over 5~Gyr of evolution,
implying that there would be no apparent connection between initially
circular orbits and chaotic diffusion over Gyr time scales. Nevertheless,
this mechanism cannot be totally ruled out, because we cannot discard
the possibility that the initial orbit of Mercury had some eccentricity.

Here, we investigate the possibility that the high eccentricity and
inclination of Mercury originated during the instability related to
the migration of the giant planets, in the framework of the so-called
``jumping Jupiter'' model. The giant planets in the Solar System
did not form on their current orbits but suffered two types of migration.
The first one was gas driven migration \citep{2001MNRAS.320L..55M},
which acted on planets during the earliest stages when the nebular
gas disk was still present in the system. The present paradigm of
this type of migration is provided by the Grand Tack model \citep{2011Natur.475..206W}.
The second one was planetesimal driven migration \citep{1984Icar...58..109F},
caused by the gravitational interaction of the giant planets with
the disk of planetesimals that remained beyond the orbit of Neptune,
once the gas has totally dissipated. Its current paradigm is given
by the Nice model \citep{2005Natur.435..459T}. According to the standard
scenario, the terrestrial planets started their formation during the
gas nebula phase and are completely formed after $\sim100$~Myr (e.g.
\citealp{1998Icar..136..304C}). Therefore, the orbital architecture
of the inner planets provides important constraints on both migration
models. For example, the terrestrial planet system is extremely sensitive
to the secular evolution of the giant planets \citep{2009A&A...507.1053B,2012ApJ...745..143A,2013MNRAS.433.3417B,2015arXiv151008448K},
and in particular, a slow and smooth migration of the outer planets
would destabilize the orbits of the inner planets.

We focus here on a variant of the Nice model, known as the jumping
Jupiter model \citep{2009A&A...507.1041M}. The present version of
this model assumes that the system of giant planets was initially
constituted of five bodies: Jupiter, Saturn, and three ice giants.
During the planetesimal driven migration, the system temporarily develops
an instability phase involving mutual close encounters between planets,
which ends up by the ejection of one of the ice giants after an encounter
with Jupiter \citep{2011ApJ...742L..22N}. This makes Jupiter jump
inwards while Saturn jumps outwards, and the period ratio between
Jupiter and Saturn changes nearly instantaneously from an initial
value of $\sim1.5$ to the current $\sim2.5$. This model has several
advantages \citep{2012AJ....144..117N} over the models of smooth
planetesimal driven migration, and has had a great success in reproducing
the dynamical properties of several minor bodies populations \citep{2013ApJ...768...45N,2014ApJ...784...22N,2014AJ....148...25D,2015AJ....150...68N,2015arXiv150106204M,2015AJ....150..186R,2016Icar...266..142B}.

The role of the giant planet evolution on the formation of the terrestrial
planets has been addressed in several recent works \citep{2003AJ....125.2692L,2011A&A...526A.126W,2013ApJ...773...65L,2014RSPTA.37230174J,2015MNRAS.453.3619I}.
Most of these studies focus on forming the terrestrial planets with
the right mass and at the right distance from the Sun, but give less
attention to their orbital eccentricities and inclinations. In some
cases, Mercury is not even taken into account in the models. Although
the field is under continuous improvement (for a review see \citealp{2014prpl.conf..595R}),
the formation of terrestrial planets is still a poorly understood
process, and the current models do not constrain well the final orbits.
Nevertheless, we know that circularization and alignment of the orbital
planes should be expected as a result of dynamical friction during
the accretion of the planetary embryos in a disk of planetesimals
\citep{2005dpps.conf...41K,2006Icar..184...39O,2008ApJ...685.1247M}.
We also know that the terrestrial planets are not expected to suffer
any migration process during or after their formation \citep{2014Icar..232..118M}.

In this paper, we present results on the evolution of the terrestrial
planets in the jumping Jupiter model, with particular focus on Mercury's
orbit. Our main assumption is that the terrestrial planets were completely
formed on nearly circular and coplanar orbits before the occurrence
of the giant planets instability. The terrestrial orbits lie initially
on the invariable plane of the giant planets. Our goal is to measure
the effects of the instability on the orbit of Mercury, and to assess
the main mechanisms that are responsible for those effects. In the
following sections, we describe the methods (sect. 2), present the
results (sect. 3), and give our conclusions (sect. 4).

\section{Methods}

We have performed a series of numerical simulations of the evolution
of the Solar System planets during the jumping Jupiter instability.
The system of giant planets is initially constituted of Jupiter, Saturn
and three Neptune-size planets, located in a mutual resonant configuration
that is the supposedly outcome of the previous gas driven migration
phase \citep{2014ApJ...795L..11P}. In particular, Jupiter and Saturn
are locked in the 3:2 mean motion resonance, with Jupiter slightly
outside of its present orbit. The initial osculating values of semimajor
axis $a$, eccentricity $e$, and inclination $I$ are shown in Table
\ref{tab:Planetary-initial-conditions}.

The five giant planets migrate according to different evolutions previously
obtained in \citet{2012AJ....144..117N}. These authors performed
realistic simulations of migration, where the giant planets interact
with a massive disk of planetesimals initially located beyond the
outermost planet. The different evolutions arise from different parameters
of the planetesimal disk that they considered. In their simulations,
\citet{2012AJ....144..117N} stored the planets' positions and velocities
in a file at 1~yr intervals over a total time span of 10~Myr. We
have not reproduced these simulations; rather we have mimicked the
migration by reading the stored positions and interpolating them using
the approach described in \citet{2013ApJ...768...45N}. The particular
cases analyzed here produce interactions with the ejected ice giant
that make the other giant planets to experience a few radial jumps.
The net inwards jump of Jupiter is $\sim0.3$~AU, and the instability
occurs about $\sim6$~Myr after the start of the simulations.

In our simulations, the terrestrial planets are initially located
at their present mean distances from the Sun, but in almost circular
and coplanar orbits. Although this might have not been the actual
case, because the early terrestrial planets orbits might have been
somewhat excited, we adopt this assumption because we want to isolate
the effects of the jumping Jupiter evolution from the effects related
to specific initial conditions. The adopted orbital values are shown
in Table \ref{tab:Planetary-initial-conditions}. The remaining orbital
elements, namely longitude of node $\Omega$, longitude of perihelion
$\varpi$, and mean longitude $\lambda$ have been chosen at random
between $0^{\circ}$ and $360^{\circ}$. For each migration case of
the giant planets, we generated 100 different sets of initial conditions
for the terrestrial planets with different values of $\Omega,\varpi,\lambda$. 

The simulations have been carried out using an hybrid version of the
SWIFT\_MVS symplectic integrator that interpolates the stored planetary
positions of the giant planets to the desired time step, and propagates
the terrestrial planets taking into account their mutual perturbations
and the perturbations from the giant planets. In this approach, the
terrestrial planets do not perturb the giant ones. Relativistic corrections
to Mercury's orbit have been introduced by an additional acceleration
term as in \citet{1991AJ....101.2287Q}. The integration time step
was 0.01~yr, and the total time span of each simulation was 10~Myr.
For a few instability cases, we have performed simulations over much
longer time spans (up to 100~Myr) that do not change our main results
and conclusions.

It is worth recalling that all the instability evolutions considered
here satisfy the constraints defined in \citet{2012AJ....144..117N},
namely, the final orbits of the outer planets are similar to the real
orbits. For example, the proper mode in Jupiter's eccentricity is
excited to its present value by the planetary encounters. All the
evolutions also satisfy the terrestrial planets constraint in that
Jupiter's orbit discontinuously evolves during planetary encounters.
This is needed to avoid secular resonances with the terrestrial planets,
which would otherwise lead to a disruption of the terrestrial planets
system \citep{2009A&A...507.1053B,2011A&A...526A.126W}.

\section{Results}

\begin{figure*}
\caption{The terrestrial planets orbits in an example of jumping Jupiter evolution.
Red dots are the final mean values of 400 fictitious orbits (100 orbits
per planet), when no relativistic correction is applied to Mercury's
orbit. Blue dots are similar results, but applying the relativistic
correction. The final values are averages over the last 1~Myr of
evolution. The initial conditions of the fictitious orbits are represented
by the small black dots at $e\simeq0$, $I\simeq0$. Open triangles
give the current mean values of the orbital elements, and error bars
give their secular variations (maximum and minimum excursions of the
elements) over 5~Myr. }

\begin{centering}
\includegraphics[width=1\textwidth]{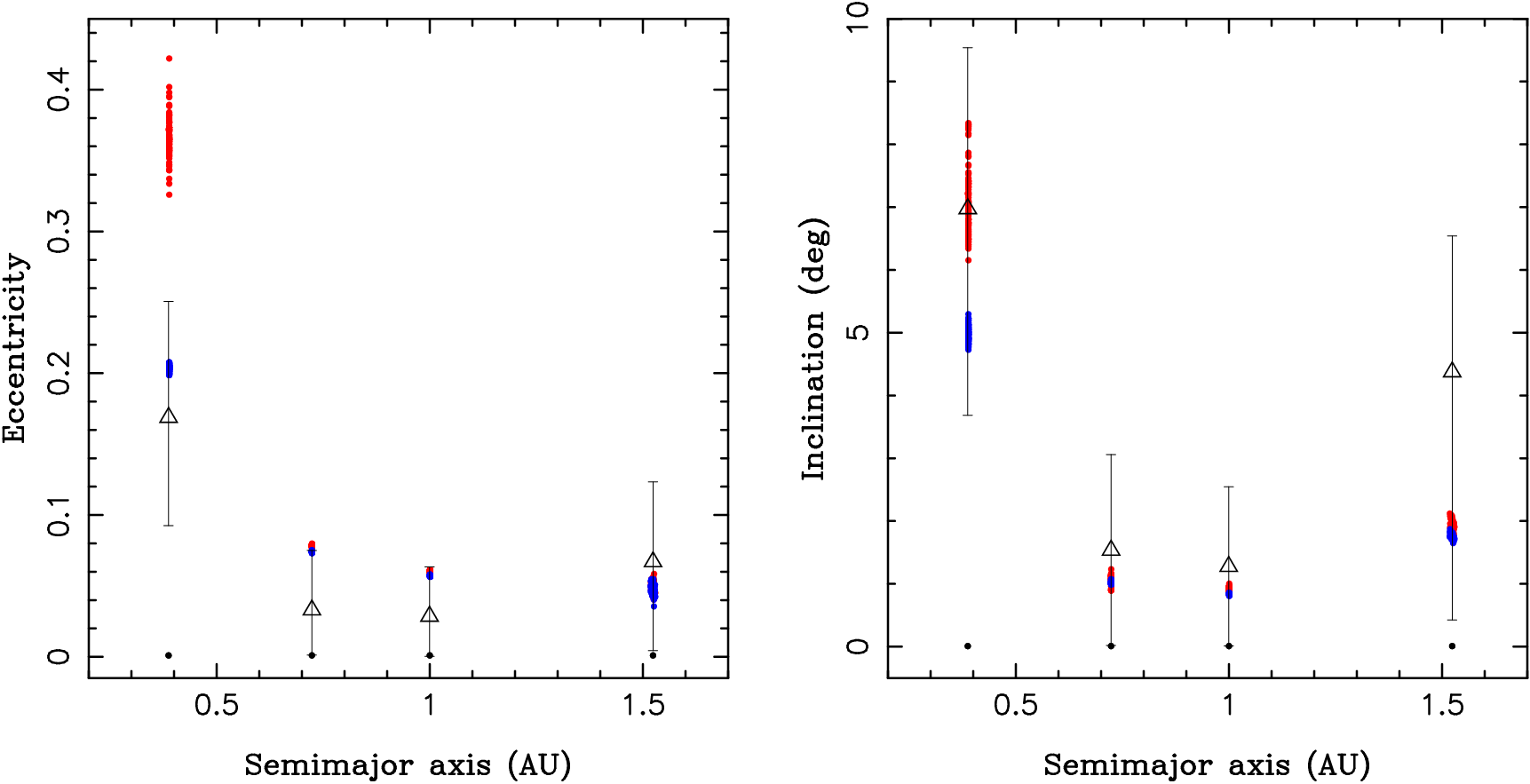}
\par\end{centering}

\label{fig1}
\end{figure*}

Figure \ref{fig1} shows the results of our simulations for one specific
case of the jumping Jupiter instability. Other instability cases produce
similar results, although those of Fig. \ref{fig1} provide best fits
to Mercury's orbit without degrading the fits to the other terrestrial
planets. Moreover, the specific instability case considered in Fig.
\ref{fig1} has also been successfully tested against the constraints
of the various minor bodies populations (\citealp{2013ApJ...768...45N};
\citealp{2014AJ....148...25D}; \citealp{2014ApJ...784...22N}; \citealp{2015AJ....150...68N};
\citealp{2015AJ....150..186R}). In the runs including the relativistic
corrections (blue dots in Fig. \ref{fig1}), Mercury's eccentricity
reached final mean values of $\sim0.2$ that are very close to the
present value of $0.17$. 

It is worth noting that in all the instability cases considered in
this study, we found that Mercury's eccentricity and inclination always
become less excited when relativistic effects are taken into account
than when these effects are ignored (cf. blue dots \textit{vs.} red
dots in Fig. \ref{fig1}). This happens because general relativity
speeds up the precession frequency $g_{1}$ of Mercury's perihelion.
Faster values of $g_{1}$ make resonances with the perihelion
frequency of Jupiter, $g_{5}$, to become less strong by reducing
their effective widths, and this leads to less excited final eccentricities.
In principle, the relativistic corrections do not introduce any direct
drift on the regression frequency $s_{1}$ of Mercury's node, but
they produce an indirect effect that also leads to less excited inclinations.
A similar result, but in a different context, has been pointed out
by \citet{2008Icar..196....1L}, who found that including relativistic
effects in the long term dynamics simulations of the Solar System
planets led to a more bounded chaotic evolution of the orbit
of Mercury over Gyr time scales.

\begin{figure}
\caption{Evolution of: (a) the eccentricity of Mercury; (b) the secular angle
$\varpi_{1}-\varpi_{5}$; (c) the secular frequencies of the perihelion
of Mercury (crosses) and Jupiter (circles); and (d) the semimajor
axis of Jupiter. The jumping Jupiter instability occurs between 5.71
and 5.74~Myr (vertical dashed line). The excitation of Mercury's
eccentricity up to $\sim0.2$ is related to a temporary trapping in
the linear secular resonance $g_{1}-g_{5}$ between the perihelia
of Mercury and Jupiter. In panel (c), the secular frequencies have
been numerically computed from the corresponding $\varpi$ time series,
over windows of 0.1~Myr in width.}

\begin{centering}
\includegraphics[width=0.5\columnwidth]{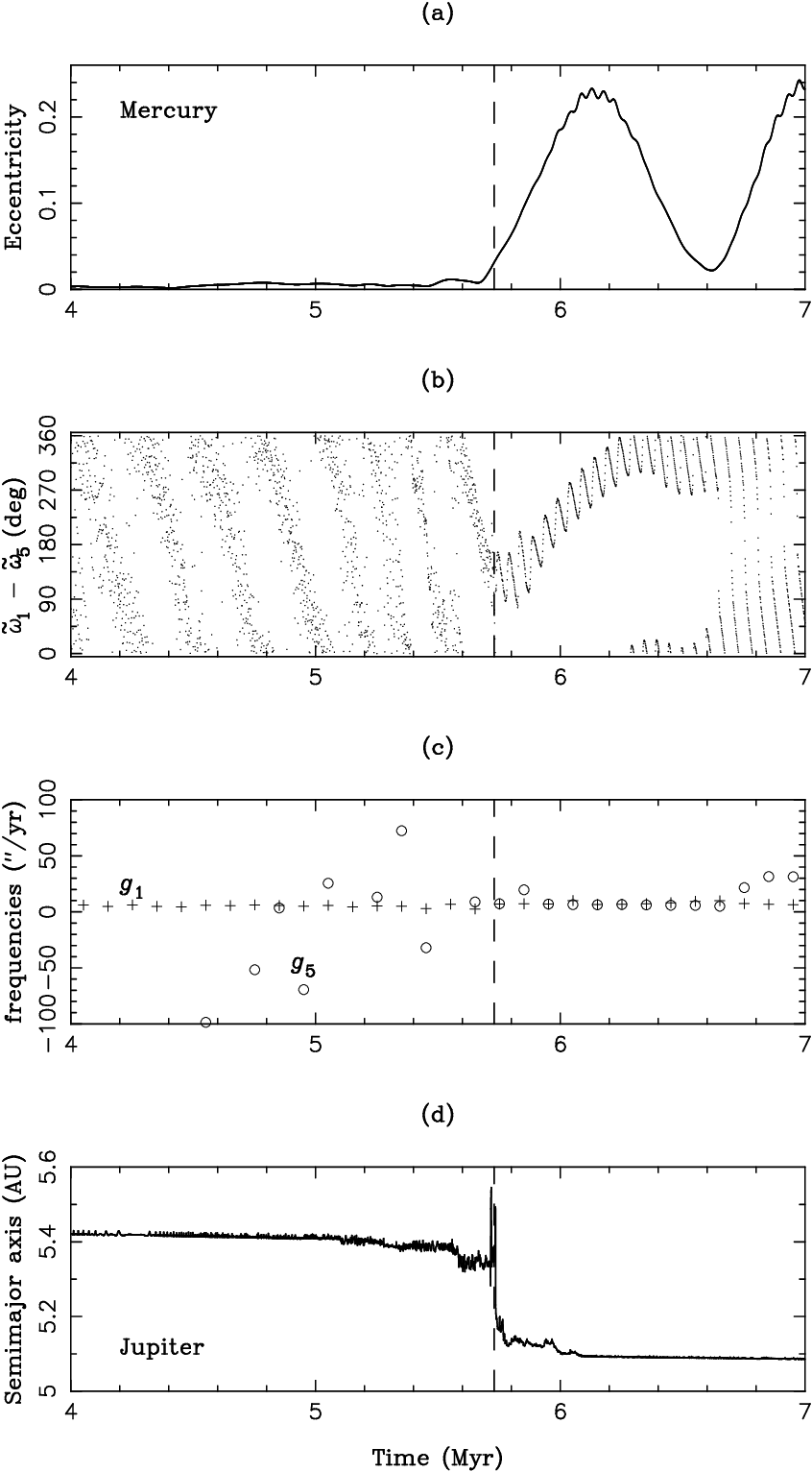}
\par\end{centering}

\label{fig-ecc}
\end{figure}

We verified that the excitation of Mercury's eccentricity is driven
by the fast sweeping of the linear secular resonance $g_{1}-g_{5}$,
in agreement with previous studies (e.g. \citealp{2009A&A...507.1053B}).
This is illustrated in Fig. \ref{fig-ecc} for one of the simulations
with relativistic corrections shown in Fig. \ref{fig1}. The raise
of Mercury's $e$ is accompanied by the libration of the angle $\varpi_{1}-\varpi_{5}$
over a short period of time. The value of $g_{5}$ shows significant
variations before the instability (Fig. \ref{fig-ecc}c), but after
Jupiter jumps (Fig. \ref{fig-ecc}d) it stabilizes and approaches
the value of $g_{1}$, driving the system into a temporary resonance
capture. We recall that permanent capture does not happen
in this context because the adiabatic threshold is broken as a consequence
of the discontinuous evolution of the giant planets.

Figure \ref{fig-inc}a shows the evolution of the inclination of Mercury
in the same simulation of Fig. \ref{fig-ecc}, together with the evolution
of the angles $\Omega_{1}-\Omega_{7}$ and $\Omega_{1}-\Omega_{2}$,
where the index $_{2}$ refers to Venus and $_{7}$ to Uranus (Fig.
\ref{fig-inc}b,c). We find that the inclination suffers two different
excitation processes, both related to secular resonances involving
the frequency of Mercury's longitude of node, $s_{1}$. The first
one is related to a temporary capture in the $s_{1}-s_{2}$ secular
resonance with Venus' node before the instability. The corresponding
resonant angle, $\Omega_{1}-\Omega_{2}$, librates around $0^{\circ}$
and the inclination is slightly excited by $\sim2^{\circ}$. The main
excitation event occurs immediately after the instability, and is
related to a temporary capture in the $s_{1}-s_{7}$ secular resonance
with Uranus' node. The corresponding resonant angle, $\Omega_{1}-\Omega_{7}$,
librates around $180^{\circ}$ and the inclination is strongly excited
up to $\sim9^{\circ}$, very similar to the present value of $7^{\circ}$.
The evolution of the secular frequencies $s_{1}$ and $s_{7}$ is
shown in Fig. \ref{fig-inc}d, together with the evolution of Uranus
semimajor axis (Fig. \ref{fig-inc}e). We can see that the jump of
Uranus from $\sim11$ to $\sim17$~AU approaches the value of $s_{7}$
to that of $s_{1}$, once again driving the system into a temporary
resonance capture. It is possible that the excitation of Mercury's
inclination actually arises from the indirect effect of the $s_{7}$
mode in the orbit of Venus (which has an important secular coupling with
Mercury; \citealp{2015ApJ...799..120B}) or Jupiter, rather than from
the direct effect of the $s_{7}$ mode in the orbit of Mercury. Unfortunately,
we cannot confirm this because the motion during and immediately after
the instability is extremely irregular, and a Fourier analysis of
the time series is useless to assess the role of the different oscillation
modes. It is worth noting, however, that during this last
excitation event, the system remains captured in the $s_{1}-s_{2}$
resonance, leading to a nodal coupling between Mercury, Venus, and
Uranus.

In general, the instability models analyzed here also provide reasonably
good fits to the other terrestrial planets. The eccentricities and
inclinations of Venus and the Earth are very well reproduced, as well
as the eccentricity of Mars. Final mean values always lie within the
range of secular variations. In a few models, the eccentricity of
Venus and the Earth are slightly more excited than required, but never
above the corresponding secular maxima. The mean inclination of Mars
is usually underestimated by some $2^{\circ}$-$3^{\circ}$, but never
below the corresponding secular minimum. 

We have also tested the final configurations using Fourier
analysis to verify if the secular architecture of the planets was
correctly reproduced. This has been carried out through a short numerical
integration of the final state of the planets, with no migration.
We found that all the secular proper frequencies are correctly reproduced.
Mercury's eccentricity and inclination are mainly contributed by the
$e_{11}$ and $I_{11}$ modes (i.e. the proper modes), respectively,
as it is in the present solar system. The other planets display the
same behavior, as expected, although some proper modes have slightly
different amplitudes, that differ from the current ones by less than
30\% in most cases, and up to a factor of two in the worst case.

\section{Conclusions}

Our results allows us to conclude that the jumping Jupiter instability
can produce the presently large values of eccentricity and inclination
of Mercury, even in the extreme case when it is assumed that this
planet formed in a circular and coplanar orbit. We tested different
instability models and some of them produces final orbits that fit
very well to the current orbit of Mercury, while also keeping a good
fit to the remaining planets, in terms of both orbital elements
and secular architecture. We found that the excitation in Mercury's
eccentricity is driven by the oscillation mode of the perihelion frequency
of Jupiter, $g_{5}$, while the excitation in inclination is driven
by oscillation mode of the node frequency of Uranus, $s_{7}$. We
also found that the relativistic correction of the perihelion frequency
of Mercury, $g_{1}$, has to be included in the model in order to
avoid excessive excitation of Mercury's eccentricity and inclination.
The jumping Jupiter evolution provides a more robust model than conservative
chaotic evolution, since it not only explains the transition of Mercury
from nearly circular and coplanar orbit to the present state, but
also fulfills several other constraints imposed by the many populations
of solar system objects.

\acknowledgements{We wish to thank an anonymous referee for his/her helpful comments
and criticism. This work has been supported by the Brazilian National
Research Council (CNPq) through fellowship 312292/2013-9 and grant
401905/2013-6 within the Science Without Borders Program, and by NASA's
Emerging Worlds program. Simulations has made use of the cluster of
the Department of Astronomy of the National Observatory of Rio de
Janeiro, acquired through CAPES grant 23038.007093/2012-13 }

\begin{table}[H]
\caption{Initial Orbital Elements and Masses of the Planets. \label{tab:Planetary-initial-conditions}}

\centering{}\medskip{}
\begin{tabular}{ccccc}
\hline 
Planet & Mass ($M_{\mathrm{Jup}}$) & $a$ (AU) & $e$ & $I$ ($^{\circ}$)\tabularnewline
\hline 
Mercury & 0.00017 & 0.387 & 0.001 & 0.01\tabularnewline
Venus & 0.00256 & 0.723 & 0.001 & 0.01\tabularnewline
Earth & 0.00318 & 1.000 & 0.001 & 0.01\tabularnewline
Mars & 0.00034 & 1.524 & 0.001 & 0.01\tabularnewline
Jupiter & 1.00000 & 5.469 & 0.003 & 0.05\tabularnewline
Saturn & 0.29943 & 7.457 & 0.011 & 0.02\tabularnewline
Ice \#1 & 0.05307 & 10.108 & 0.017 & 0.11\tabularnewline
Ice \#2 & 0.05307 & 16.080 & 0.006 & 0.07\tabularnewline
Ice \#3 & 0.05411 & 22.172 & 0.002 & 0.05\tabularnewline
\hline 
\end{tabular}
\end{table}

\begin{figure}
\caption{Evolution of: (a) the inclination of Mercury; (b) the secular angle
$\Omega_{1}-\Omega_{7}$; (c) the secular angle $\Omega_{1}-\Omega_{2}$;
(d) the secular frequencies of the nodes of Mercury (crosses) and
Uranus (circles); and (e) the semimajor axis of Uranus. The jumping
Jupiter instability is indicated by the vertical dashed line. The
excitation of Mercury's inclination up to $\sim9^{\circ}$ is related
to a temporary trapping in the linear secular resonance $s_{1}-s_{7}$
between the nodes of Mercury and Uranus. Before the instability, a
slight excitation up to $2^{\circ}$ is related to a secular resonance
with the node of Venus. In panel (d), the secular frequencies have
been numerically computed from the corresponding $\Omega$ time series,
over windows of 0.1~Myr in width.}

\label{fig-inc}
\end{figure}

\setcounter{figure}{2}
\begin{figure}
\caption{}

\begin{centering}
\includegraphics[width=0.5\columnwidth]{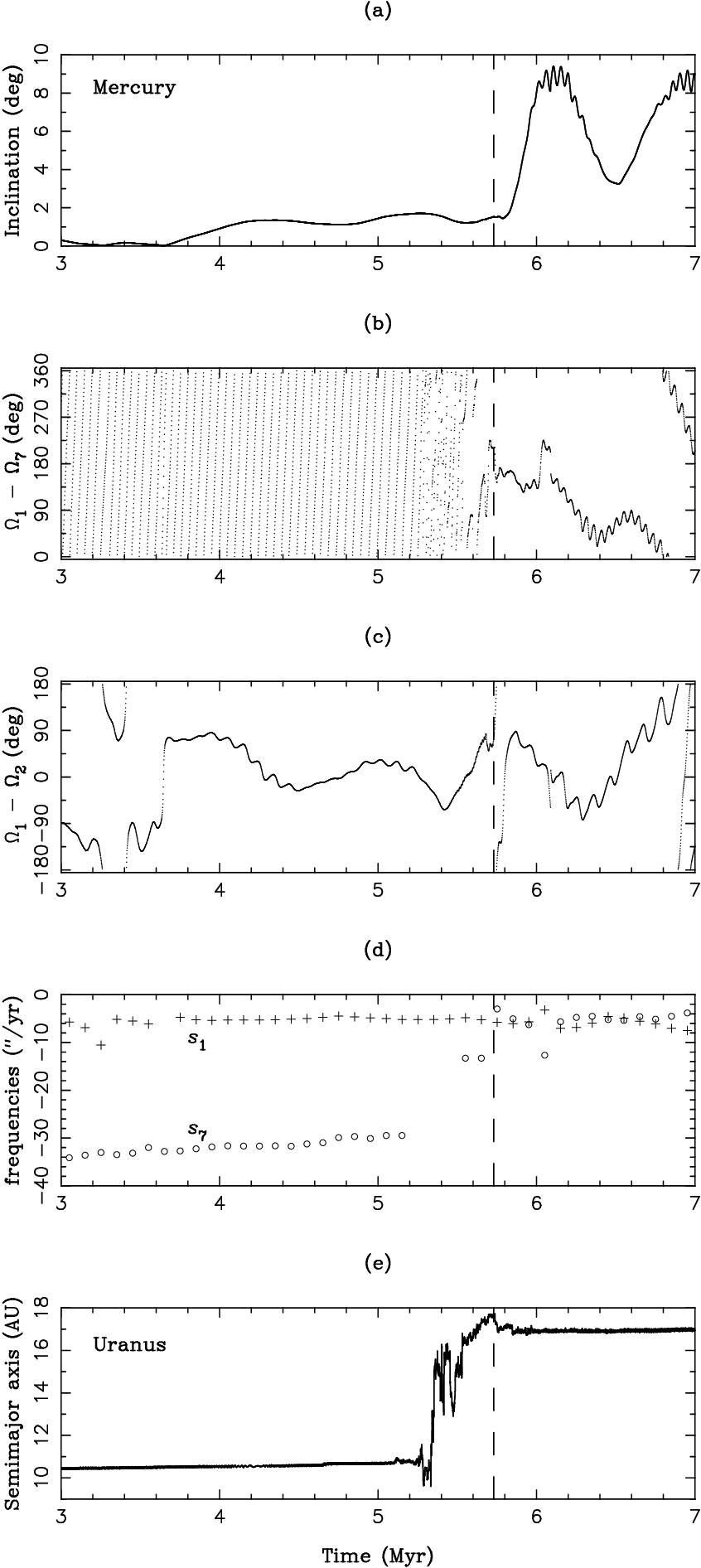}
\par\end{centering}

\end{figure}

\end{document}